\documentclass[11pt]{article}
\usepackage{moriond2000,epsfig}

\def\MNRAS{\textit{ Mon.\ Not.\ Roy.\ Astr.\ Soc.\ }}
\def\ApJ{\textit{ Astroph.\ Journ.\ }}

\def\AA{\textit{ Astron.\ Astroph.\ }}

\def\be{\begin{equation}}
\def\ee{\end{equation}}
\def\bea{\begin{eqnarray}}
\def\eea{\end{eqnarray}}

\begin{document}

\title{Gravitational Lensing and Cluster of Galaxies}

\author{Jean-Paul KNEIB}

\address{Laboratoire d'Astrophysique, Observatoire Midi-Pyr\'en\'ees\\
14 avenue E.-Belin, F-31400 Toulouse, France}

\maketitle\abstracts{
\begin{quotation}
  {\cal \it Multiple images, giant arcs, Einstein cross, fold, cusp,
    lip, caustics, critical lines, sources, mapping, time delay,
    arclets, weak shear, magnification bias, depletion, ellipticities,
    polarization, smearing, convergence, kernel, mass reconstruction,
    sub-structure, dark halos, natural telescope, amplification,
    distant galaxies, faintest sources ...}
\end{quotation}
This is a non exhaustive list of favorite terms used when using
gravitational lensing in clusters of galaxies.  I will introduce here
lensing in clusters as a useful tool in modern {\it observational}
cosmology.  I will give a summary of what we are learning in terms of
cluster {\it mass distribution} in the strong and weak regime, and
what information we will gain in terms of the cluster physics. I will
underline the benefit of using cluster-lenses as {\it natural
  telescopes} to probe the distant Universe.  }

\section{Introduction}
  
\begin{figure}
\centerline{\psfig{file=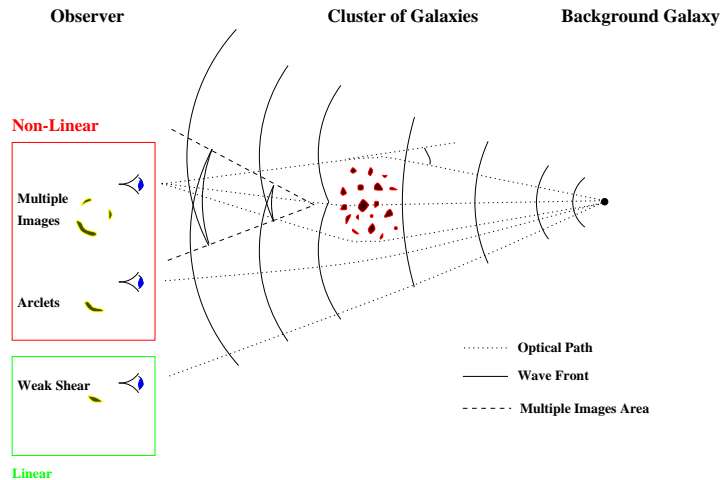,width=0.6\textwidth}}
\caption{Gravitational lensing in clusters: A simple representation of
  how gravitational images are formed.  }
\end{figure}

Cluster of galaxies are the largest and most massive bounded
structures in the Universe. Due to their important mass density they
locally deformed the Space-Time.  Therefore, the wave front of the
light coming from any distant galaxy (or more generally any emitting
light source) passing through a galaxy cluster will be distorted.
Moreover, for the most massive clusters the mass density in the core is
sufficiently high to break the wave front into pieces hence producing
multiple images of distant galaxies, which usually form these
extraordinary gravitational giant arcs (the {\it strong lensing}
domain). Distant galaxies will thus appeared distorted and magnified,
we usually call them arclets because of their noticeable elongated
shape tangentially aligned toward the cluster center. Note however
that their shape is a combination of the intrinsic shape and the
distortion induced by the cluster, thus a lensed galaxy can appear
round if its intrinsic orientation is perpendicular to the shear
direction.  If the alignment between the observer, the cluster and 
distant galaxies is less perfect the distortion induced by the cluster
will be less important and can not be recognize immediately --
statistical method are required -- (we entering the {\it weak regime}
domain).  Indeed in this region, the shape of the galaxies are
dominated by their intrinsic ellipticity or worse contaminated by the
distortion of the camera and/or the point spread function (PSF) of the
image.

In the thin lens approximation (which usually holds for
cluster-lenses {\it e.g} Schneider, Ehlers \& Falco, 1992), the
deflection of light between the position of the source ${\vec
  \theta_S}$ and the position of the image ${\vec \theta_I}$ is given
by the lensing mapping equation:
\begin{equation}
{\vec \theta_S} = {\vec \theta_I} - 
   {2 {\cal D} \over c^2}   {\vec \nabla}\phi^{2D}_N({\vec \theta_I})
= {\vec \theta_I} - {\vec \nabla}\varphi({\vec \theta_I})
\end{equation}
where ${\cal D} = D_{LS}/D_{OS}$ is the angular distance ratio between
the Lens and the Source and the Observer and the Source [this ratio
therefore depends on the redshift of the cluster $z_L$ and the
background source $z_S$, as well as - but only weakly - on the
cosmological parameter $\Omega_m$ and $\Omega_\lambda$], and
$\phi^{2D}_N$ is the Newtonian projected gravitational potential, and $\varphi$
the lensing potential.
This transformation is thus a mapping from Source plane to Image plane,
and the Hessian of this transformation relates a source element of the Image
to the Source plane:
\begin{equation}
{{d\vec \theta_S} \over {d\vec \theta_I}}= {\cal H} = {\cal A}^{-1} =
\pmatrix{
1 - \partial_{xx} \varphi & - \partial_{xy} \varphi\cr
- \partial_{xy} \varphi & 1 - \partial_{yy} \varphi \cr
}
=
\pmatrix{
1 - \kappa - \gamma_1 & -\gamma_2 \cr
-\gamma_2 & 1 - \kappa + \gamma_1 \cr
}
\end{equation}
where we have defined the convergence $\kappa=\Sigma/2\Sigma_{crit}$,
 the shear $\vec\gamma = (\gamma_1,\gamma_2)$ and the magnification
matrix ${\cal A}$. This matrix also governs the shape transformation 
from the Source to the Image plane.

Thus, cluster lenses can be used in 2 ways: {\it i)} Firstly by
understanding and modeling the gravitational optics of this system:
by probing the {\bf total mass} distribution of the cluster -- which
explains the observed image configuration and distortions --, by
constraining the {\bf distance} of the lensed galaxies -- the more
distant the more distorted they are --, to put constraints on the
cosmological parameters - although this is a second order effect --.
{\it ii)} Secondly as a {\it Natural telescope}: galaxies seen through
massive cluster cores are amplified by the gravitational lensing
effect making them easier to study in details; the faintest sources --
which would otherwise remain unknown -- can be detected/identified as
the sensitivity of instruments is boosted by the gravitational
amplification.

\section{Cluster Lens Properties}

\subsection{ Strong Lensing}

Massive clusters can produce multiple images, this will happen when
the surface mass density of the cluster reach or is larger than the
{\it critical density} $\Sigma_{crit}= {c^2\over 4G}{D_{OS} D_{OL}
  \over D_{LS} }$. The configuration of multiple images tells us about
the structure of the mass distribution. A cluster with one dominant
clump of mass will produce {\it fold} or {\it cusp} arcs, radial arcs
({\it e.g} MS2137.3-2353: Fort et al 1992, Mellier et al 1993; 
AC114: Natarajan et al 1998;
A383: Smith et al 2000); a bimodal cluster can produce straight
arcs ({\it e.g} Cl2236-04: Kneib et al 1994a), triplets (A370: Kneib et
al 1993, Bezecourt et al 1999) or even triangular image; a very
complex structure with lots of massive halos in the core can produce
multiple image system with seven or more images of the same source
({\it e.g} Cl2244-04, A2218).

A particular useful and popular mass estimate in the strong lensing
regime is the mass within the Einstein radius $R_E$:
$M(<R_{E}) = \pi\Sigma_{crit} \theta_E^2$;
$R_E$ is the location of the critical line for a circular mass
distribution, usually approximated by the arc radius
$R_{arc}$. It is a very handy expression -- independent of the mass
profile for a circular mass symmetry --, but one should be careful in
using it: either because the arc used to derive the mass as a unknown
redshift, or the arc is a single image and thus does not trace the
Einstein radius (for a singular isothermal sphere model, a single
image can not be closer to twice the Einstein radius or it will have a
counter image!), or the mass distribution is very complex with a lot
of sub-structure. In conclusion, {\bf this estimator does generally
  overestimate the mass}.

The only route to accurately constrain the mass in cluster cores is
to use multiple images with preferably a spectroscopic redshift to
absolutely calibrate the mass. As the problem is generally degenerate
--{\it in the sense that there is not a single mass distribution but a
  family of model that is fitting the observables --}, one should used
physically motivated representation of the mass distribution and
adjust it in order to best reproduce the different family of multiple
images ({\it e.g} Kneib et al 1996). As the position of the images are
known to great accuracy and are usually located in different places of
the cluster cores a simple mass model with one clump can usually not
reproduce the image configuration. The lens model needs to include the
cluster galaxies to match up the image configuration and positions. As
there is not an infinite number of multiple images and thus number of
constraints, it is important to limit the number of free parameters of
the model and keep it physically motivated -- as in the end -- we are
interested to derive physical properties of the cluster. Alternative
method, using non-parametric description have been explored ({\it
  e.g.}  Abdelsalam et al 1999), but usually lack the resolution of a
parametric form due to the large dynamical range of the mass density
expected in a cluster core - but clearly this is an interesting
approach than should explored further.

The strong lensing mass modeling technique is an iterative method, in
the sense that once a multiple images is securely identified, other
multiple images systems can be discovered using morphological
or color criteria as well as the predictions from the lens model. 
The lens model can then predict redshift for these
multiple systems (Kneib et al 1993, Natarajan et al 1998) as well as
for the arclets (Kneib et al 1994b, 1996): on the basis that on average
a distant galaxy is randomly orientated, and its ellipticity follow a
relatively peaked ellipticity distribution. These prediction can then
be tested/verified ({\it e.g.} Ebbels et al 1998) and an improved mass
model can be derived integrating the new constraints.

The ultimate step of strong lensing modeling is to constrain the
cosmological parameters. This can be undertaken, when in a cluster core,
a sufficient number of multiple images ($>3$) are identified and for
which spectroscopic information can be measured (see Golse et al
 this conference).

\subsection{ Weak Lensing}

In the weak lensing regime the game is different: we measure the {\it
  mean} ellipticity and/or the {\it mean} number density of faint
galaxies, and we want to relate these statistics to the mean surface
mass density $\kappa$ of the cluster. There are two issues in doing that:\\
$\bullet$ one for a {\it theorist}: What is the best method to reconstruct the
mass distribution $\kappa$ (as a mass map or a radial mass profile)
from the `shear field' $\vec\gamma$ and/or the magnification bias?\\
$\bullet$ one for an {\it observer}: How best determined the `true'
ellipticity of a faint galaxy which is smeared by a PSF barely smaller
than the object (when using ground-based images) that is not circular
(camera distortion, tracking errors ...) and not stable in time?
How best estimate the variation in the number density of faint
galaxies due to the lensing effect, taking into account the crowding
effect due to the cluster and the intrinsic fluctuations in the
distribution of galaxies?\\

Various approaches have been proposed to solve these two problems, and
we can distinguish two families of methods: {\bf direct} and {\bf
  inverse} methods.

For the theorist issue, the direct approaches are: {\it i)} the Kaiser
\& Squires (93) method (an integral method, that express $\kappa$ as
the convolution of $\vec\gamma$ by a kernel) and subsequent
refinements ({\it e.g.} Seitz et al. 1995, 1996); {\it ii)} the local
inversion method (Kaiser, 1995, Schneider, 1995, Lombardi \& Bertin
1998) integrates the gradient of $\vec \gamma$ within the boundary of
the observed field to then derive $\kappa$.  The inverse approach
works on $\kappa$ or the lensing potential $\varphi$ and uses maximum
likelihood (Bartelman et al 1996, Schneider et al 2000) or maximum
entropy method (Bridle et al 1998) to determine the most likely mass
distribution (as a 2D mass map or a 1D mass profile) that reproduce
the shear field $\vec\gamma$ and/or the variation in the faint galaxy
number densities. These inverse methods are of great interest as they
allow: to quantify the errors in the resultant mass maps or mass
estimates (Bridle et al 1999), as well as to introduce external
constraints (such as strong lensing, or X-ray).

For the observer, before any data handling, the first priority is to
choose the telescope that will minimize the source of noise in the
determination of the ellipticity of faint galaxies. Although the {\it
  Hubble Space Telescope} ({\it HST}) has the best characteristics in
terms of the PSF, it has a very limited field of view
not really appropriate to probe the large scale distribution of a
cluster (note this is of course less of a problem when looking at high
redshift clusters). What is really needed is a wide field imager and
excellent seeing conditions! \\
Then, we can use a direct approach using for example the Kaiser,
Squires and Broadhurst (1995) method [KSB implemented in the {\it
  imcat} software], or any other improvement of it (Luppino \& Kaiser
1998, Rhodes et al 2000, Kaiser 2000): that relates the true
ellipticity to the observed ellipticity correcting it from the
smearing of an
elliptical PSF (using the second moments of the galaxy and the PSF).\\
The inverse approach use maximum likelihood method to find the source
galaxy shape that when convolved by the local PSF reproduce best the
observed galaxy ({\it e.g.} Kuijken, 2000). Again the inverse approach
has the advantage to give directly an uncertainty in the parameter
recovery.
The weak-shear mass reconstruction techniques have been applied to
wide-field camera data (UH8k, CFH12k, ESO-WFI, CTIO-MegaCam)
and impressive results have started to be published
on a high redshift super-cluster (Kaiser et al 1998) and on low
($z<0.1$) redshift clusters (Joffre et al 2000). For high ($z>0.5$) redshift
clusters large aperture telescope ({\it e.g.} Clowe et al 2000) or
{\it HST} (Hoekstra et al 2000) are probably more adequate.

\subsection{ Cluster Galaxies Halos}

We know that galaxies are massive and that their stellar content does
only represent a small part of their total mass. Although the
existence of a dark halo has been obvious very early for disk galaxies
with the study of their flat velocity curve out to large radius ({\it e.g.}
van Albada et al 1985),
the existence of a dark halo has been accepted for ellipticals
relatively recently ({\it e.g.} Kochanek 1995, Rix et al 1997). These studies
found that the stellar content dominates the central part of the
galaxies, but at distance larger than the effective radius
the dark halo dominates the total mass.

Galaxy lensing effect were first detected in clusters by Kassiola et
al (1992) who notes that lengths of the triple arc in Cl0024+1654 can
only be explained if the galaxies near the B image were massive
enough. Detailed treatment of the galaxy contribution to the cluster mass
became important with the refurbishment of the {\it HST}
as first shown by Kneib et al 1996 -- who concluded that galaxies (and
their dark halos) in cluster cores contributes by about 10\% of the
total mass. The theory of what is usually called galaxy-galaxy lensing
in clusters was first discussed in details by Natarajan \& Kneib 1997,
and application to data followed shortly (Natarajan et al 1998 and Geiger \&
Schneider 1998).  A recent analysis of this effect in various
cluster-lenses at various redshift seems to indicate an increase of
the cluster ellipticals dark halo size with redshift (Natarajan et al
2000). These new developments are very interesting, as for the first
time they offer a powerful tool to relate the total mass of cluster
galaxies to their morphological aspects. This tool will probably help
us in better understanding the strong morphological evolution seen in
cluster galaxies.
The standard direct {\it weak shear} methods generally miss the
small scale fluctuations (typically the galaxy halo scales) because of
the {\it averaging} of the galaxy ellipticities. Thus dedicated
methods are necessary to probe this effect in the weak shear method.
The only easy route is to use an inverse approach which will examine
the galaxies {\bf individually}.

\subsection{ Lensing and other Estimators}

Gravitational lensing allow to measure the {\it total} mass
distribution of clusters -- and this without making any assumption on
the cluster physical state. Other estimators always require some
assumption when trying to relate the observables to the {\it total}
mass.  Generally these assumptions looks reasonable but may suffer
strong bias due to the unknown physical state of the cluster.  By
providing the {\it total} mass, lensing does constitute a {\bf key}
tool to understand cluster physics. Probably then, the best way is to
first derive the total lensing mass using lensing, and then from other
observations derive physical properties of the cluster like: dynamical
parameters for the galaxy velocities (Natarajan \& Kneib 1996), the
temperature profile of the X-ray gas (Pierre et al 1996), the baryon 
fraction or the equilibrium status of the cluster -- however lensing mass 
estimates have also their limitations (in particular line of sight projection
effects).

The alternative way is to compare the different estimators
directly.  As an example, X-ray mass estimates generally differ
sensibly from the strong/weak lensing estimates - however not always.
The differences can be explained for different reasons depending on
the cluster studied ({\it e.g.} Miralda \& Babul 1995): {\it i)}
projection effects: 2 clusters can be aligned on the line of sight and
boost the lensing mass; {\it ii)} simple X-ray modeling: for example
multiphase gas distribution are necessary in cooling flow clusters
({\it e.g} Allen, Fabian \& Kneib 1996); {\it iii)} non-thermal effect
can modify the central mass estimates; {\it iv)} the cluster just
suffer a major merger event and the dynamical state of the gas can not
be considered as in thermal equilibrium.  

The canonical lensing
clusters Abell 370 and Cl0024+1654 are two examples were the X-ray
mass and lensing mass do not agree.  For Abell 370, the disagreement is
directly visible on the ROSAT/HRI X-ray surface brightness map that
only peaks on the Southern cD galaxy, despite the lensing mass model
requires a bimodal structure with equivalent mass around the 2 cDs --
this difference, may however disappear when better X-ray observations
(with {\it Chandra} and {\it XMM-Newton}) are made of this cluster.  
For Cl0024+1654, the
X-ray emission is weak compared to the large Einstein ring observed. A
recent redshift survey of $\sim$300 cluster galaxies (Czoske et al
2000) does however unveils some of the mystery. The redshift
histogramme show a complex structure with a main relaxed structure
compatible with the X-ray emission and a foreground structure along
the line of sight that boost the lensing strength of this cluster.

Recently, Sunyaev-Zeldovich (SZ) effect has been routinely measured on
the most X-ray luminous clusters ({\it e.g.} Carlstrom et al 2000).
As SZ is probing the intra-cluster gas in a different way than X-ray
observations, it is important to use SZ as a complementary approach to the
lensing, X-ray and galaxies velocities estimators as a detailed study
will teach us a lot on the cluster physics. Attempts of combining these
different informations were presented during this Conference.

Ideally,  one wants to looks at the mass properties of clusters as a
function of time to derive their evolution. But to do that we need
well defined samples of clusters, studied in an homogenous way.
However, precise and systematic comparison is still relatively rare in
the literature, as they usually rely on published data, either on the
X-ray or on the lensing part, and thus do no tackle a well defined
sample, nor do they address carefully the limits and bias of the two
different approaches.  This is however currently changing rapidly, as
a number of dedicated surveys based on well defined cluster catalogue
(Ebeling et al 1996, 2000) are in progress ({\it e.g.} Czoske et al,
 this conference).

\subsection{Dark lenses?}

It has been known for a while that some of the multiple image quasars
can only be modeled if an important external shear contribution was
added to the main lens contribution (Keeton et al 1997). In other
cases the image separation between the multiple quasars is so large
that large M/L for the main lensing galaxy are required. Thus, the
existence of  so-called {\it dark clusters} has been discussed.  Recent deep
inspection of these systems, followed by optical spectroscopy seems
to reveal that {\it dark clusters} are not so dark after all
(Benitez et al 1999, Kneib et al 2000, Soucail et al 2000).  A
systematic deep survey of those multiple quasar systems where either a
too high M/L ratio or a large external shear is required would be
useful to understand whether {\it dark or not so dark} lump of matter
are affecting the lensing of the quasars.

In this respect the detection of a dark lump of matter near the
cluster Abell 1942 by Erben et al (2000) is very puzzling.
Either it is an extremely rare (?) cosmic conspiracy in the
distribution of the faint galaxy ellipticities or what is detected is
really a massive dark concentration of mass which true nature should
be understood.

\section{Cluster Lenses as Nature Telescopes}

The most massive clusters can be used as efficient {\it Natural
Telescope}.  The key feature of
these systems is that any distant object seen through the clusters is
amplified (and distorted).  This amplification can easily exceed a
factor of $>2$--$3\times$ for the central 4 sq.\ arcmin of the lens and
will be still higher than $>1.4\times$ over a 20 sq.\ arcmin field of view
for the most massive clusters.
The amplification provides a magnified view of a correspondingly
smaller region of the source plane -- so the 4 sq.\ arcmin region seen
through the core of a cluster lens will actually translate to a $<2$
sq.\ arcmin patch of the background sky.  Thus a lens provides a more
sensitive, but also more restricted, view of the high redshift sky.
These effects, the amplification and the reduction in the available
area, compensate each other for a source population with a count
slope, $\alpha=1$, where $N(>S) \propto S^{-\alpha}$ (or equivalently for
a count slope $\gamma=0.4$, where $N(m) \propto 10^{\gamma m}$).
 However, the sources we identify in the lens field
will be on average intrinsically $\sim 2\times$ fainter than those
identified in an equivalent blank field.

Depending on the waveband used, we will either see, more, or less,
sources than in blank field regions. As a first example, in the sub-mm
waveband $\alpha$ is indeed very close to unity at the faintest flux
(Blain et al.\ 1999) and so we expect to detect equivalent numbers of
sources in lensed fields as in a blank field in the {\it same}
exposure time (see also Ivison, this conference).  In
the optical and Near-Infra red (NIR), the slope $\gamma$ is about
0.3  at the faintest flux, thus we expect less sources than in
blank field. Finally, in the Mid-Infra red (MIR), the slope $\alpha$
is $\sim 1.5$ at the faintest flux (Metcalfe et al 1999)
and we detect more sources than in the blank fields. A particular,
the faintest MIR sources were detected in the deep ISOCAM pointing of
Abell 2390 (Altieri et al 1999).

The cluster-lens technique therefore allows you to reach below the
sensitivity limit of normal observations.  To successfully employ this
lens technique we need to be able to correct the observations for the
amplification by the cluster using a detailed mass model of the lens
constructed from {\it HST} imaging is necessary (see section 2.1).

This technique has three major advantages: {\it i)} the image
resolution in the source plane is effectively finer leading: to a
fainter confusion limit for the sub-mm maps and MIR observations, and
to smaller resolution elements in optical/NIR allowing to better
identify the morphological aspects of these faint sources; {\it ii)}
cluster-lenses are some of the best studied regions of the
extragalactic sky -- thus deep multi-wavelength observations are
generally available making the identification of distant galaxy much
easier; {\it iii)} in the case of rare events where the amplification
is larger than 10, detailed physical observation of the distant lensed
galaxy can be made on morphological aspects (Pell\'o et al 1999,
Soucail et al 1999) or on spectroscopical aspects (Pettini et al
2000).

Similar lensing techniques are starting to be used to search for
high-redshift supernovae [SN] ({\it e.g.} Sullivan et al 2000) or to
detect Lyman-$\alpha$ emitters (Ellis et al 2000, in prep). In the
case of a detection of a SN in a multiple image, if we are able to
measure a time delay, it will give a unique way to precisely constrain
the {\it Hubble} parameter $H_0$.

\section{Future and Prospects}

Since the discovery of giant arcs and arclets in the end of the 80's
gravitational lensing applications in cluster of galaxies have grown
considerably.\\
$\bullet$
We are now able to reconstruct the mass distribution in clusters in
great details from the galaxy scale to the virial radius. The lensing
mass estimate will be usefully compared to other mass estimators to
provide critical information on the cluster physics (from the largest
cluster scale to the galaxy scale) on well defined cluster samples.\\
$\bullet$
Wide field survey of {\it mass selected} cluster using lensing
techniques will allow to make direct comparison to analytic/numerical
models of the Universe and thus better understand the growth of
structure and the large scale distribution of mass. It will also
confirm or otherwise the existence of dark lump of mass.\\
$\bullet$
Multiple images in cluster cores are about to measure directly the
cosmological parameters through an optical geometrical test of the
curvature of the Universe, although more spectroscopic and mass
modeling are needed, it is a very clean way to tackle this problem.\\
$\bullet$
Likewise, time dependant phenomenom like Supernovae or AGN
fluctuations if observed behind well-known lensing clusters, may prove
to be a very accurate way to probe the $H_0$, as it
has been initiated using multiple quasars.\\
$\bullet$
Finally, massive clusters will always be the {\it unique place} to
look at to boost telescope and instrument sensitivities to push ahead the
discoveries to the faintest detection level.

\section*{Acknowledgments}
{\small\it
I acknowledge support from the LENSNET network as well as from
INSU/CNRS.  I thanks all my collaborators for the fruitful work we are
conducting to better understand our Universe using lensing as a
powerful tool. I specially thanks Yannick Mellier who chaired the
organization of this wonderful meeting and who first with Bernard Fort
taught me about lensing in clusters.
}

\end{document}